\documentclass[10pt,journal,doublecolumn]{IEEEtran}
\usepackage{amsmath,amssymb}
\usepackage[dvips]{graphicx}
\usepackage{amsfonts}
\usepackage{latexsym}

\usepackage{graphicx,picture,calc}

\usepackage[format=hang]{subfig}
\newtheorem{theorem}{\textit{{Theorem}}}
\newtheorem{lemma}{{\textit{{Lemma}}}}
\newtheorem{corollary}{{\textit{{Corollary}}}}

\newtheorem{remark}{\textit{{Remark}}}

\newtheorem{problem}{\textit{{Problem}}}

\begin{document}
\title{Asymptotically Locally Optimal Weight Vector Design for a Tighter Correlation Lower Bound of Quasi-Complementary Sequence Sets}
\author{Zilong~Liu,~
         Yong~Liang~Guan,~
         Wai~Ho~Mow~
\thanks{Zilong Liu and Yong Liang Guan are with the School
of Electrical and Electronic Engineering, Nanyang Technological
University, Singapore. E-mail: \{zilongliu, eylguan\}@ntu.edu.sg.}
\thanks{Wai Ho Mow is with the Department of  Electronic and Computer Engineering, Hong Kong University of Science
and Technology, Clear Water Bay, Hong Kong S.A.R., China. E-mail: w.mow@ieee.org.}
}
 \maketitle

 \begin{abstract}
A quasi-complementary sequence set (QCSS) refers to a set of two-dimensional matrices with low non-trivial aperiodic auto- and cross- correlation sums. For multicarrier code-division multiple-access applications, the availability of large QCSSs with low correlation sums is desirable. The generalized Levenshtein bound (GLB) is a lower bound on the maximum aperiodic correlation sum of QCSSs. The bounding expression of GLB is a fractional quadratic function of a weight vector $\mathbf{w}$ and is expressed in terms of three additional parameters associated with QCSS: the set size $K$, the number of channels $M$, and the sequence length $N$. It is known that a tighter GLB (compared to the Welch bound) is possible only if the condition $M\geq2$ and $K\geq \overline{K}+1$, where $\overline{K}$ is a certain function of $M$ and $N$, is satisfied. A challenging research problem is to determine if there exists a weight vector which gives rise to a tighter GLB for \textit{all} (not just \textit{some}) $K\geq \overline{K}+1$ and $M\geq2$, especially for large $N$, i.e., the condition is {asymptotically} both necessary and sufficient. To achieve this, we \textit{analytically} optimize the GLB which is (in general) non-convex as the numerator term is an indefinite quadratic function of the weight vector. Our key idea is to apply the frequency domain decomposition of the circulant matrix (in the numerator term) to convert the non-convex problem into a convex one. Following this optimization approach, we derive a new weight vector meeting the aforementioned objective and prove that it is a local minimizer of the GLB under certain conditions.


\end{abstract}


\section{Introduction}

In recent years, multicarrier code-division multiple-access (MC-CDMA) based on the quasi-/perfect- complementary sequence set (in abbreviation, QCSS/PCSS) has attracted much attention due to its potential to achieve low-/zero- interference multiuser performance \cite{HHC2001}, \cite{HHCBook}. Here, a QCSS (or PCSS) refers to a set of two-dimensional matrices with low (or zero) non-trivial auto- and cross- correlation sums \cite{Liu-TCOM2014}$-$\cite{LiuGuanMow14}. In this paper, a complementary sequence is also called a complementary matrix, and vice versa.

To deploy a QCSS (or PCSS) in an MC-CDMA system, every data symbol of a specific user is spread by a complementary matrix by simultaneously sending out all of its row sequences over a number of non-interfering subcarrier channels. Because of this, the number of row sequences of a complementary matrix, denoted by $M$, is also called the \textit{number of channels}. At a matched-filter based receiver, de-spreading operations are performed separately in each subcarrier channel, followed by summing the correlator outputs of all the subcarrier channels to attain a correlation sum which will be used for detection.

A PCSS may also be called a mutually orthogonal complementary sequence set (MOCSS) \cite{TSENG72}$-$\cite{Han-2011}, a concept extended from mutually orthogonal Golay complementary pairs (GCPs) \cite{Golay61}$-$\cite{LiuLiGuan-2013}. However, a drawback of PCSS is its small set size \cite{HHC08-REAL}. Specifically, the set size (denoted by $K$) of PCSS is upper bounded by the number of channels, i.e., $K\leq M$. This means that a PCSS based MC-CDMA system with $M$ subcarriers can support at most $M$ users only. Against such a backdrop, there have been two approaches aiming to provide a larger set size, i.e., $K>M$. The first approach is to design zero- or low- correlation zone (ZCZ/LCZ) based complementary sequence sets, called ZCZ-CSS \cite{Fan07}, \cite{HHC08} or LCZ-CSS \cite{Liu-TCOM2011}. A ZCZ-CSS (LCZ-CSS) based MC-CDMA system is capable of achieving zero- (low-) interference performance but requires a closed-control loop to dynamically adjust the timings of all users such that the received signals can be quasi-synchronously aligned within the ZCZ (LCZ). A second approach is to design QCSS which has uniformly low correlation sums over all non-trivial time-shifts. As such, QCSS can be utilized to achieve low-interference performance with a simpler timing-control system. To the authors' best knowledge, the first aperiodic correlation lower bound of QCSS was derived by Welch in \cite{Welch74}, which states:
\begin{equation}\label{Welch_bound_for_cc}
 \delta^2_{\max} \geq M^2N^2\frac{\frac{K}{M}-1}{K(2N-1)-1},
\end{equation}
where every quasi-complementary sequence is a matrix of order $M\times N$ (thus, every row sequence has length of $N$) with assumed energy of $M^2N^2$. The aforementioned set size upper bound of PCSS, namely, $K\leq M$, can also be obtained from (\ref{Welch_bound_for_cc}) by setting $\delta_{\max}=0$. On the other hand, if $0<\delta_{\max}\ll MN$, one can show that $K>M$, meaning that a larger set size can be supported by QCSS.

Recently, a generalized Levenshtein bound (GLB) for QCSS has been derived by Liu, Guan and Mow in [\ref{LiuGuanMow14}, \textit{Theorem 1}]. The key idea behind the GLB (including the Levenshtein bound \cite{Levenshtein99}) is that the weighted mean square aperiodic correlation of any sequence subset over the complex roots of unity should be equal to or greater than that of the whole set which includes all possible complex roots-of-unity sequences. The Levenshtein bound was extended from binary sequences to complex roots-of-unity sequences by Bozta\c{s} \cite{Serdar1998}. A lower bound for aperiodic LCZ sequence sets was derived in \cite{Peng04} by an approach similar to Levenshtein's.

In its bounding equation, GLB is a function of the ``simplex" weight vector $\mathbf{w}$, the set size $K$, the number of channels $M$, and the row sequence length $N$.  
A necessary condition (shown in [\ref{LiuGuanMow14}, \textit{Theorem 2}]) for the GLB to be tighter than the Welch bound is that $K\geq \overline{K}+1$,
where
\begin{equation}\label{nece_cond_QCSSBd2}
\overline{{K}}\triangleq \left \lfloor
4(MN-1)N\sin^2\frac{\pi}{2(2N-1)} \right \rfloor,
\end{equation}
with
\begin{equation}\label{nece_cond_QCSSBd2_}
\lim\limits_{N\rightarrow \infty}\overline{K}=\left \lfloor \frac{\pi^2M}{4} \right \rfloor.
\end{equation}
Although a ``step-function" weight vector was adopted in [\ref{LiuGuanMow14}, (34)], it only leads to a tighter GLB for $K\geq3M+1$. As a matter of fact, the tightness of GLB remains unknown for
\begin{displaymath}
\frac{\left \lfloor \frac{\pi^2M}{4} \right \rfloor}{M}<\frac{K}{M}<3+\frac{1}{M},
\end{displaymath}
when $N$ is sufficiently large.

The main objective of this paper is to optimize and then tighten the GLB for \textit{all} $K\geq \overline{K}+1$ (instead of \textit{some}). For this, we are to find a (locally) optimal weight vector which is used in the bounding equation. A similar research problem was raised in \cite{Levenshtein99} for traditional binary sequences (i.e., non-QCSS with $M=1$). See \cite{LiuParaGuanBozas14} for more details. The optimization of GLB on QCSS (with $M\geq2$), however, is more challenging because an analytical solution to a non-convex GLB (in terms of weight vector $\mathbf{w}$) for \textit{all} possible cases of $(K,M)$ is in general intractable. 

We first adopt a frequency-domain optimization approach in Section III-B to minimize the (non-convex) fractional quadratic function of GLB. This is achieved by properly exploiting the specific structure of the circulant quadratic matrix in the numerator of the fractional quadratic term of GLB. Following this optimization approach, we find a new weight vector which leads to a tighter GLB for \textit{all} $(K,M)$ cases satisfying $K\geq \overline{K}+1$ and $M\geq2$, asymptotically (in $N$). Our finding shows that the condition of $K\geq \overline{K}+1$, shown in [\ref{LiuGuanMow14}, Theorem 2], is not only necessary but also sufficient, as $N$ tends to infinity. Moreover, in Section III-C, it is proved that the newly found weight vector is a local minimizer to the fractional quadratic function of GLB, asymptotically.

We then examine in Sections IV two weight vectors which were presented in \cite{LiuParaGuanBozas14} for the tightening of the Levenshtein bound on conventional single-channel (i.e., $M=1$) sequence sets. We extend their tightening capability to GLB on multi-channel (i.e., $M\geq2$) QCSS, although the proof is not straightforward. It is shown that each of these two weight vectors gives rise to a tighter GLB (over the Welch bound) for several small values of $M$ provided that $K\geq \overline{K}+1$. It is also noted that the GLB from the newly found weight vector is (in general) tighter than the GLBs from these two (earlier found) weight vectors, as shown by some numerical results.


\section{Preliminaries}
In this section, we first present some necessary notations and define QCSS. Then, we give a brief review of GLB.
\subsection{Introduction to QCSS}
For two complex-valued sequences $\mathbf{a}=[a_0,a_1,\cdots,a_{N-1}]$ and $\mathbf{b}=[b_0,b_1,\cdots,b_{N-1}]$, their aperiodic correlation function at time-shift $\tau$ is defined as
\begin{equation}
\rho_{\mathbf{a},\mathbf{b}}(\tau)=\left
\{
\begin{array} {c@{\quad \quad}l}
\sum\limits_{t=0}^{N-1-\tau} a_{t}  b^{*}_{t+\tau} , & 0{\leq}\tau{\leq}(N-1); \\
\sum\limits_{t=0}^{N-1+\tau} a_{t-\tau} b^*_{t}, & -(N-1){\leq}\tau{\leq}-1;\\
0, & |\tau|\geq N.
\end{array} \right.
\end{equation}
When $\mathbf{a}\neq\mathbf{b}$, $\rho_{\mathbf{a},\mathbf{b}}(\tau)$
is called the aperiodic cross-correlation function (ACCF); otherwise, it is called the aperiodic auto-correlation function (AACF). For simplicity, the AACF of $\mathbf{a}$ is denoted by $\rho_{\mathbf{a}}(\tau)$.

Let $\mathcal{C}=\{\mathbf{C}^0,\mathbf{C}^1,\cdots,\mathbf{C}^{K-1}\}$ be a set of $K$ matrices, each of order $M\times N$ (where $M\geq 2$), i.e.,
\begin{equation}\label{matrix_form_of_CC}
\begin{split}
\mathbf{C}^\nu  =\left [ \begin{array}{c} \mathbf{c}^\nu_0\\ \mathbf{c}^\nu_1 \\ \vdots \\
\mathbf{c}^\nu_{M-1}
\end{array} \right ]_{M\times N} =\left [ \begin{matrix}
c^\nu_{0,0} & c^\nu_{0,1} & \cdots & c^\nu_{0,N-1}\\
c^\nu_{1,0} & c^\nu_{1,1} & \cdots & c^\nu_{1,N-1}\\
\vdots      & \vdots      & \ddots & \vdots\\
c^\nu_{M-1,0} & c^\nu_{M-1,1} & \cdots & c^\nu_{M-1,N-1}\\
\end{matrix} \right ],
\end{split}
\end{equation}
where $0\leq \nu \leq K-1$. Define the ``aperiodic correlation sum" of matrices $\mathbf{C}^\mu$ and $\mathbf{C}^\nu$ as follows,
\begin{equation}\label{aperiodic_corr_CC}
{\rho}_{\mathbf{C}^\mu,\mathbf{C}^\nu}(\tau)=\sum\limits_{m=0}^{M-1}\rho_{\mathbf{c}^\mu_m,\mathbf{c}^\nu_m}(\tau),~~0\leq \mu,\nu\leq K-1.
\end{equation}
Also, define the aperiodic auto-correlation tolerance $\delta_{a}$
and the aperiodic cross-correlation tolerance $\delta_{c}$ of $\mathcal{C}$
as
\begin{displaymath}
\begin{array}{l}
\delta_{\text{a}}\triangleq\max \left \{\Bigl |\rho_{\mathbf{C}^\mu,\mathbf{C}^\mu}(\tau) \Bigl |:~\begin{matrix}
0<\tau\leq N-1,\\
0\leq \mu\leq K-1.
\end{matrix}
\right \},\\
\delta_{\text{c}}\triangleq\max \left \{\Bigl |\rho_{\mathbf{C}^\mu,\mathbf{C}^\nu}(\tau) \Bigl |:~\begin{matrix}
0\leq \tau \leq N-1,\\
\mu\neq\nu, 0\leq \mu,\nu\leq K-1.
\end{matrix}
 \right \}\\
\end{array}
\end{displaymath}
respectively. Moreover, define the aperiodic tolerance (also called the ``maximum aperiodic
correlation magnitude") of $\mathcal{C}$ as $\delta_{\max}\triangleq\max\{\delta_{\text{a}},\delta_{\text{c}}\}$. When
$\delta_{\max}=0$, $\mathcal{C}$ is called a \textit{perfect complementary sequence set} (PCSS); otherwise, it is called a \textit{quasi-complementary
sequence set} (QCSS)\footnote{QCSS can also be defined with respect to the ``periodic correlation sums". The interested reader may refer to \cite{Liu-WCL13}.}. In particular, when $M=2$ and $K=1$, a PCSS reduces to a matrix consisting of two row sequences which have zero out-of-phase aperiodic autocorrelation sums. Such matrices are called Golay complementary matrices (GCMs) or Golay complementary pairs (GCPs) in this paper, and either sequence in a GCP is called a Golay sequence.

Note that the transmission of a PCSS or a QCSS requires a multi-channel system. Specifically, every matrix in a PCSS (or a QCSS) needs $M\geq2$ non-interfering channels for the separate transmission of $M$ row sequences. This is different from the traditional single-channel sequences with $M=1$ only.

\subsection{Review of GLB}
Let $\mathbf{w}=[w_0,w_1,\cdots,w_{2N-2}]^{\text{T}}$ be a ``simplex" weight vector which is constrained by
\begin{equation}\label{Leven_weight_vector}
w_i\geq0,~~i=0,1,\cdots,2N-2, ~~\text{and} ~~
\sum\limits_{i=0}^{2N-2}w_i=1.
\end{equation}
Define a quadratic function 
\begin{equation}\label{Leven-quad-fun-equ}
\begin{split}
Q(\mathbf{w},a) & \triangleq\mathbf{w}^\text{T} \mathbf{Q}_{a}
\mathbf{w}\\
 & =a\sum\limits_{i=0}^{2N-2}w^2_i+\sum\limits_{s,t=0}^{2N-2}\tau_{s,t,N}w_s
w_t,
\end{split}
\end{equation}
where $\mathbf{Q}_{a}$ is a $(2N-1)\times(2N-1)$ circulant matrix with all of its diagonal
entries equal to $a$, and its off-diagonal entries $\mathbf{Q}_{a}(s,t)=\tau_{s,t,N}$, where $s\neq t$ and
\begin{equation}\label{defi_of_l}
0\leq \tau_{s,t,N}\triangleq\min\left \{|t-s|,2N-1-|t-s| \right \} \leq N-1.
\end{equation}

The GLB for QCSS over complex roots of unity in \cite{LiuGuanMow14} is shown below.
\vspace{0.1in}
\begin{lemma}\label{generalized_welch_bound_for_cc}
\begin{equation}\label{generalized_welch_bound_for_cc-equ}
 \delta^2_{\max}  \geq M \left [
N-\frac{{Q}\left(\mathbf{{w}},\frac{N(MN-1)}{K}\right)}{1-\frac{1}{K}\sum\limits_{i=0}^{2N-2}w_i^2}
\right ].
\end{equation}
A weaker simplified version of (\ref{generalized_welch_bound_for_cc-equ}) is given below.
\begin{equation}\label{simplified_GLB}
 \delta^2_{\max}  \geq M \left [
N-{{Q}\left(\mathbf{{w}},\frac{MN^2}{K}\right)}
\right ].
\end{equation}
\end{lemma}

%

\vspace{0.1in}

\begin{remark}
Setting ${\mathbf{w}}=\frac{1}{2N-1}(1,1,\cdots,1)$, the GLB reduces to the Welch bound for QCSS in (\ref{Welch_bound_for_cc}).
\end{remark}

\vspace{0.1in}

\begin{remark}\label{rmk_nece_cond}

[\ref{LiuGuanMow14}, \textit{Theorem 2}] For the GLB to be tighter than  the corresponding Welch bound, it is \textit{necessary} that $K\geq \overline{K}+1$, where $\overline{K}$ is defined in (\ref{nece_cond_QCSSBd2}).
\end{remark}

%

\vspace{0.1in}

\begin{remark}

[\ref{LiuGuanMow14}, \textit{Corollary 1}] Applying the weight vector ${{\mathbf{{w}}}}$ with
\begin{equation}\label{leven_LCZ_weighting_vector}
 {{w}}_i=\left \{
\begin{array}{cl}
 \frac{1}{m}, & ~~i\in \{0,1,\cdots,m-1\};\\
 0,   & ~~i\in \{m,m+1,\cdots,2N-2\};
\end{array}
\right .
\end{equation}
where $1\leq m \leq N$, to (\ref{generalized_welch_bound_for_cc-equ}), we have
\begin{equation}\label{ZL_corollary_4_equ}
 \delta^2_{\max}  \geq \max_{1\leq m\leq N}
\frac{3MNKm-3M^2N^2-MK(m^2-1)}{3(mK-1)}. 
\end{equation}
The lower bound in (\ref{ZL_corollary_4_equ}) is tighter than the Welch bound for QCSS
in (\ref{Welch_bound_for_cc}) if one of the two following conditions is fulfilled:

(1): $3M+1\leq K \leq 4M-1$, $M\geq2$ and
\begin{equation}\label{N_range_for_K_leq_4M}
N\geq \left \lfloor \frac{K-1+\sqrt{-3K^2+(12M-6)K+12M+1}}{2(K-3M)}
\right \rfloor+1;
\end{equation}

(2): $K\geq4M$, $M\geq2$ and $N\geq2$.
\end{remark}

\section{Proposed Weight Vector for Tighter GLB }

\subsection{Motivation}
\begin{figure}[!ht]
\centering
\scalebox{0.55}
{\includegraphics{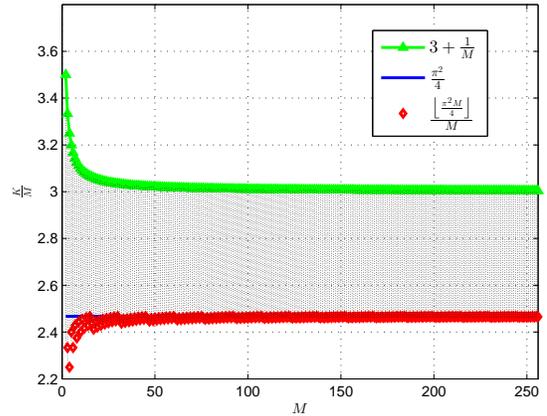}}
\caption{The tightness of GLB is unknown over the shaded $K/M$ zone, where $N$ is sufficiently large.}
\label{Fig_GLBgap}
\end{figure}

The necessary condition in \textit{Remark \ref{rmk_nece_cond}} implies that for a given $M,N$, the Welch bound for QCSS cannot be improved if $K \leq \overline{{K}}$, where $\overline{{K}}$ is defined in (\ref{nece_cond_QCSSBd2}). On the other hand, the weight vector in (\ref{leven_LCZ_weighting_vector}) can only lead to a tighter GLB for $K\geq3M+1$. Because of this,
the tightness of GLB is unknown in the following ambiguous zone.
\begin{equation}\label{GLBgap_}
\frac{\overline{K}}{M} < \frac{K}{M} < 3+\frac{1}{M}.
\end{equation}
For sufficiently large $N$, the above $K/M$ zone further reduces to
\begin{equation}\label{GLBgap}
\frac{\left \lfloor \frac{\pi^2M}{4} \right \rfloor}{M}<\frac{K}{M} < 3+\frac{1}{M},
\end{equation}
by recalling (\ref{nece_cond_QCSSBd2_}). One may visualize this zone in the shaded area of Fig. \ref{Fig_GLBgap} for $2\leq M \leq 256$.

We are therefore interested in finding a weight vector which is capable of optimizing and tightening the GLB for \textit{all} (rather than \textit{some}) $K\geq \overline{K}+1$. Relating this objective to Fig. \ref{Fig_GLBgap}, such a weight vector can give us a tighter GLB for the largest $K/M$ region right above the red diamond symbols.

However, the optimization of GLB in (\ref{generalized_welch_bound_for_cc-equ}) is challenging because its fractional quadratic term (in terms of $\mathbf{w}$) is indefinite. More specifically, the quadratic term ${Q}\left(\mathbf{{w}},\frac{N(MN-1)}{K}\right)$ in the numerator is indefinite as some eigenvalues of the corresponding circulant matrix are negative when $K\geq \overline{K}+1$ [\ref{LiuGuanMow14}, Appendix B]. It is noted that indefinite quadratic programming (QP) is NP-hard \cite{MK1987}, even it has one negative eigenvalue only \cite{PV1991}. Moreover, checking local optimality of a feasible point in constrained QP is also NP-hard \cite{PS1998}. Although some optimality conditions for constrained QP have been derived by Bomze from the copositivity perspective \cite{Bomze1997-1,Bomze1997-2,Bomze2015}, the situation becomes more complicated when indefinite fractional quadratic programming (FQP) problems are dealt with. According to \cite{ABJ2014}, GLB may be classified as a standard FQP (StFQP) as the feasible set is the standard simplex. To the best of the authors' knowledge, Preisig pioneered an iterative algorithm for which convergence to a KKT point (but cannot be guaranteed to be a local minimizer) of the StFQP can be proved \cite{Preisig1996}. Two algorithms for StFQP based on semidefinite programming (SDP) relaxations are presented in \cite{ABJ2014}, yet the optimalities of the resultant solutions are unknown. As a matter of fact, the algorithms developed in \cite{ABJ2014,Preisig1996} may only be feasible for medium-scaled StFQP with $N\leq 200$. In contrast, we target at an analytical solution (as opposed to a numerical solution) which is applicable to large scale of GLB (e.g., the sequence length $N>1000$). Thus, the techniques used in \cite{ABJ2014,Preisig1996} may not be useful for the specific StFQP problem considered in this paper.

In the sequel, we introduce a frequency-domain optimization approach which finds a local minimizer (i.e., a weight vector) of the GLB. We show that the obtained weight vector leads to a tighter GLB for \textit{all} $K\geq \overline{K}+1$ and $M\geq 2$, asymptotically.

\vspace{0.1in}

\subsection{GLB from Weight Vector 1}
To tighten the GLB in (\ref{generalized_welch_bound_for_cc-equ}), we adopt a novel optimization approach in this subsection, motivated by the observation that any circulant matrix [e.g., $\mathbf{Q}_a$ in (\ref{Leven-quad-fun-equ}) which forms a part of the GLB quadratic function in (\ref{generalized_welch_bound_for_cc-equ})] can be decomposed in the frequency domain.

Define $\xi_L=\exp\left ( {-\sqrt{-1}2\pi}/{L} \right )$ and the $L$-point discrete Fourier transform (DFT) matrix as
\begin{equation}
\mathbf{F}_{L}=[f_{m,n}]_{m,n=0}^{L-1},~\text{where}~f_{m,n}=\xi^{mn}_L.
 \end{equation}
 Denote by $\mathbf{q}$ the first column vector of $\mathbf{Q}_{a}$ in (\ref{Leven-quad-fun-equ}), i.e.,
 \begin{equation}
 \mathbf{q}=[a,1,2,\cdots,N-1,N-1,\cdots,2,1]^{\text{T}}.
 \end{equation}
Let
\begin{equation}
\mathbf{v}  =\mathbf{F}_{2N-1}\mathbf{w}=[v_0,v_1,\cdots,v_{2N-2}]^{\text{T}}.
\end{equation}
It is noted that $v_0=\sum_{i=0}^{2N-2}w_i=1$. By \cite{Gray-book}, the circulant matrix $\mathbf{Q}_a$ defined in (\ref{Leven-quad-fun-equ}) can be expressed as
\begin{equation}\label{Qa_in_FreqDomain}
\mathbf{Q}_a = \frac{1}{2N-1} \mathbf{F}^{\text{H}}_{2N-1}\text{diag}({\text{\mbox{\boldmath{$\lambda$}}}}) \mathbf{F}_{2N-1},
\end{equation}
where
\begin{equation}
{\text{\mbox{\boldmath{$\lambda$}}}}  =\mathbf{F}_{2N-1}\mathbf{q}=[\lambda_0,\lambda_1,\cdots,\lambda_{2N-2}]^{\text{T}},
\end{equation}
and $\text{diag}({\text{\mbox{\boldmath{$\lambda$}}}})$ is the matrix with ${\text{\mbox{\boldmath{$\lambda$}}}}$ being the diagonal vector and zero for all the non-diagonal matrix entries. Consequently [\ref{NYY2014}, Theorem 3.1],
\begin{equation}\label{GLB_quadra_fd}
Q(\mathbf{w},a) = \frac{1}{2N-1} \sum\limits_{l=0}^{2N-2}\lambda_l \left | v_l\right |^2.
\end{equation}
Similarly,
\begin{equation}\label{GLB_quadra_fd_}
\sum\limits_{i=0}^{2N-2}w^2_i = \frac{1}{2N-1} \sum\limits_{l=0}^{2N-2}\left | v_l\right |^2.
\end{equation}
By [\ref{LiuGuanMow14}, Appendix B], we have
\begin{equation}
\lambda_0=a+(N-1)N,
\end{equation}
and
\begin{equation}\label{lambda_i_equ}
\lambda_l=a-\frac{1-(-1)^l \cos{\frac{\pi
l}{2N-1}}}{2\sin^2{\frac{\pi l}{2N-1}}},
\end{equation}
for $l=1,\cdots,2N-2$. Note that $\lambda_l=\lambda_{2N-1-l}$ for $l\in \{1,2,\cdots,N-1\}$. Moreover, we remark that
\begin{equation}\label{lambda_l1_equ}
\lambda_l>\lambda_1,~~\text{for}~2\leq l \leq N-1.
\end{equation}
This is because
\begin{enumerate}
\item if $l$ is odd:
\begin{equation}
\lambda_l-\lambda_1=\frac{\sin^2{\frac{\pi l}{2(2N-1)}}-\sin^2{\frac{\pi}{2(2N-1)}}}{4\sin^2{\frac{\pi}{2(2N-1)}}\sin^2{\frac{\pi l}{2(2N-1)}}}>0.
\end{equation}
\item if $l$ is even:
\end{enumerate}
\begin{equation}
\lambda_l-\lambda_1=\frac{\cos \frac{\pi l}{2N-1}+\cos \frac{\pi}{2N-1}}{8\sin^2{\frac{\pi}{2(2N-1)}}\cos^2{\frac{\pi l}{2(2N-1)}}}>0.
\end{equation}
To maximize the GLB in (\ref{generalized_welch_bound_for_cc-equ}), it is equivalent to consider the following optimization problem.
\vspace{0.1in}

\begin{problem}
\begin{equation}\label{opti_freqd}
\begin{split}
 & \min_{\mathbf{v}}\frac{\lambda_0 + \sum\limits_{l=1}^{2N-2}\lambda_l |v_l|^2}{2N-1-\frac{1}{K}-\frac{1}{K}\sum\limits_{l=1}^{2N-2}|v_l|^2},\\
 & ~~\text{subject to}~\mathbf{w}  =\frac{1}{2N-1}\mathbf{F}^{\text{H}}_{2N-1}\mathbf{v}\geq 0.
\end{split}
\end{equation}
\end{problem}
\vspace{0.1in}

Since $\mathbf{w}$ is real-valued, $\mathbf{v}$ is conjugate symmetric, i.e., $v_l=v^*_{2N-1-l}$ for $l=1,2,\cdots,2N-2$. Having this in mind, we define
\begin{equation}
r^2= \sum_{l=1}^{N-1}|v_l|^2=\sum_{l=N}^{2N-2}|v_l|^2.
\end{equation}
Taking advantage of the fact that $\lambda_1=\lambda_{2N-2}$ are strictly smaller than other $\lambda_l$'s with nonzero $l$ as shown in (\ref{lambda_l1_equ}), we have
\begin{equation}\label{opti_freqd_ineq}
\sum\limits_{l=1}^{2N-2}\lambda_l |v_l|^2=2\lambda_1 r^2+\sum\limits_{l=2}^{2N-3}(\lambda_l-\lambda_1) |v_l|^2\geq 2\lambda_1 r^2,
\end{equation}
where the equality is achieved if and only if $v_l=0$ for $l=2,3,\cdots,2N-3$.
Inspired by this observation, we relax the non-negativity constraint on $\mathbf{w}$, i.e., some negative $w_i$'s may be allowed (but the sum of all elements of $\mathbf{w}$ must still be equal to 1). With this, the optimization problem in (\ref{opti_freqd}) can be translated to
\begin{equation}\label{opti_freqd_trans}
\begin{split}
 & \min_{r} \min_{\sum_{l=1}^{2N-2}|v_l|^2=2r^2} \frac{\lambda_0 + \sum\limits_{l=1}^{2N-2}\lambda_l |v_l|^2}{2N-1-\frac{1+2r^2}{K}},\\
= & \min_{r} \frac{\lambda_0 + 2 \lambda_1 r^2}{2N-1-\frac{1}{K}-\frac{2r^2}{K}},
\end{split}
\end{equation}
where
 \begin{equation}\label{lambdas}
 \begin{split}
 \lambda_0 & = \frac{N(MN-1)}{K}+N(N-1),\\
 \lambda_1 & = \frac{N(MN-1)}{K}-\frac{1}{4\sin^2 \frac{\pi}{2(2N-1)}}.
 \end{split}
 \end{equation}
From now on, we adopt the setting of
\begin{displaymath}
\begin{split}
v_1 & =v^*_{2N-2}=r\exp\left (\sqrt{-1}\theta \right ),\\
v_l & =0,~\text{for}~l=2,3,\cdots,2N-3,
\end{split}
\end{displaymath}
where $r,\theta$ denote the magnitude and phase of $v_1$, respectively. Since $\mathbf{w}  =\frac{1}{2N-1}\mathbf{F}^{\text{H}}_{2N-1}\mathbf{v}$, we have
\begin{equation}\label{wtvec3_equ}
\begin{split}
\mathbf{w}&=\frac{1}{2N-1}\left  [1+2r\cos \theta, 1+2r\cos \left (\theta+\frac{2\pi}{2N-1}\right ) \right.,\\
          & ~~~~~~~~~~~~~~~~~~~~~~~~\left. \cdots,1+2r\cos \left (\theta+\frac{2\pi(2N-2)}{2N-1} \right ) \right ]^{\text{T}}.
\end{split}
\end{equation}

To optimize the fractional function in (\ref{opti_freqd_trans}), we have the following lemma.
\vspace{0.1in}

\begin{lemma}\label{lem_monofun}
The fractional function $\frac{\lambda_0 + 2 \lambda_1 r^2}{2N-1-\frac{1}{K}-\frac{2r^2}{K}}$ in terms of $r^2$ in (\ref{opti_freqd_trans}) is
\begin{enumerate}
\item[] Case 1: monotonically decreasing in $r^2$ if $K\geq\overline{K}+1$ and $\frac{\lambda_0}{|\lambda_1|}<(2N-1)K-1$;
\item[] Case 2: monotonically increasing in $r^2$ if $K\leq \overline{K}$, or $K\geq \overline{K}+1$ and $\frac{\lambda_0}{|\lambda_1|}\geq (2N-1)K-1$.
\end{enumerate}
\end{lemma}
\vspace{0.1in}

\begin{proof}
 To prove Case 1, we first show that $\lambda_1<0$ if and only if
 \begin{displaymath}
 K\geq \overline{K}+1=\left \lfloor
4(MN-1)N\sin^2\frac{\pi}{2(2N-1)} \right \rfloor+1,
 \end{displaymath}
 where $\overline{K}$ is defined in (\ref{nece_cond_QCSSBd2}). For ease of analysis, we write
  \begin{equation}\label{K_split_equ}
 4(MN-1)N\sin^2\frac{\pi}{2(2N-1)}=n+\epsilon,
  \end{equation}
where $n$ is a positive integer and $0\leq \epsilon<1$.
Thus, $\overline{K}+1=n+1$. Consequently, we have
\begin{equation}
\begin{split}
\lambda_1 & = a-\frac{1}{4\sin^2 \frac{\pi}{2(2N-1)}}\\
          & = \frac{N(MN-1)}{K}-\frac{1}{4\sin^2 \frac{\pi}{2(2N-1)}}\\
          & =\frac{N(MN-1)}{K} \left [1- \frac{K}{4N(MN-1)\sin^2 \frac{\pi}{2(2N-1)}} \right]\\
          & \leq \frac{N(MN-1)}{K} \left ( 1- \frac{n+1}{n+\epsilon} \right )\\
          & <0,
\end{split}
\end{equation}
with which the proof of Case 1 follows. The proof of Case 2 can be easily obtained by following a similar argument.
\end{proof}

\vspace{0.1in}
For Case 2 of \textit{Lemma \ref{lem_monofun}}, it can be readily shown that the minimum of the fractional function $\frac{\lambda_0 + 2 \lambda_1 r^2}{2N-1-\frac{1}{K}-\frac{2r^2}{K}}$ in (\ref{opti_freqd_trans}) is achieved at $r=0$.
Thus, the weight vector in (\ref{wtvec3_equ}) reduces to
\begin{equation}\label{wtvec3_equ2}
\begin{split}
\mathbf{w}&=\frac{1}{2N-1}\cdot \left [1, 1,\cdots,1 \right ]^{\text{T}},
\end{split}
\end{equation}
where the corresponding GLB reduces to the Welch bound in (\ref{Welch_bound_for_cc}).

Next, let us focus on the application of Case 1 for GLB tightening. In this case, we wish to know the upper bound of $r^2$ in order to minimize the fractional function of $r^2$ in (\ref{opti_freqd_trans}).
Coming back to the constraint of $\mathbf{w}$ given in (\ref{Leven_weight_vector}), $r$ and $\theta$ should satisfy
\begin{equation}
1+2r\min_{i=0,1,\cdots,2N-2}\cos \left (\theta+\frac{2\pi i}{2N-1}\right )\geq 0.
\end{equation}
Thus,
\begin{equation}
0\leq r\leq \max\limits_{\theta}\frac{-1}{\min\limits_{i=0,1,\cdots,2N-2}\cos \left (\theta+\frac{2\pi i}{2N-1}\right )}=\frac{1}{2\cos\left ( \frac{\pi}{2N-1}\right )},
\end{equation}
where the upper bound is achieved with equality when $\theta=\frac{2\pi j}{2N-1}$ for any integer $j$. By substituting $r=\frac{1}{2\cos\left ( \frac{\pi}{2N-1}\right )}$ into (\ref{wtvec3_equ}), we obtain the following weight vector.
\begin{equation}\label{wgtvec3_equ}
w_i=\frac{1}{2N-1}\left ( 1+\frac{\cos \frac{2\pi(i+j)}{2N-1}}{\cos \frac{\pi}{2N-1}}\right ),
\end{equation}
where $i=0,1,\cdots,2N-2$ and $j$ is any integer. The resultant GLB from this weight vector is shown in the following lemma. 
\vspace{0.1in}

\begin{lemma}\label{coro_GLB_from_wetvec3}
For $K\geq\overline{K}+1$ and $\frac{\lambda_0}{|\lambda_1|}<(2N-1)K-1$, we have
\begin{equation}\label{coro_GLB_from_wetvec3-equ}
\delta^2_{\max}\geq   M \left [N- \frac{K \left (\lambda_0-\frac{|\lambda_1|}{2\cos^2 \frac{\pi}{2N-1}} \right )}{(2N-1)K-1-\frac{1}{2\cos^2 \frac{\pi}{2N-1}}}\right ],
\end{equation}
where $\lambda_0,\lambda_1$ are given in (\ref{lambdas}).
\end{lemma}

\vspace{0.1in}

To analyze the asymptotic tightness of the lower bound in (\ref{coro_GLB_from_wetvec3-equ}), we note that when $N$ is sufficiently large, the second condition in \textit{Lemma \ref{coro_GLB_from_wetvec3}}, i.e.,
\begin{equation}\label{cond_Case1}
\frac{\lambda_0}{|\lambda_1|}<(2N-1)K-1,
\end{equation}
is true for $K\geq \overline{K}+1$.
To show this, we substitute $\lambda_0,\lambda_1$ into (\ref{cond_Case1}). After some manipulations, one can see that the inequality in (\ref{cond_Case1}) holds if and only if
\begin{equation}\label{cond_Case1_2}
\begin{split}
K> & 4(MN-1)N\sin^2\frac{\pi}{2(2N-1)}\\
   & +\frac{4N(N-1)\sin^2 \frac{\pi}{2(2N-1)}+1}{2N-1}.
\end{split}
\end{equation}
Carrying on the expression in (\ref{K_split_equ}), we require
\begin{equation}\label{cond_Case1_3}
K-n\geq 1>\epsilon+\frac{4N(N-1)\sin^2 \frac{\pi}{2(2N-1)}+1}{2N-1},
\end{equation}
which is guaranteed to hold for sufficiently large $N$ because $\epsilon$ is strictly smaller than 1 by assumption.
Furthermore, we note that
\begin{equation}
\begin{split}
\lim\limits_{N \rightarrow\infty}\frac{K\lambda_0}{N} & = (K+M)N,\\
\lim\limits_{N \rightarrow\infty}\frac{K|\lambda_1|}{2N\cos^2 \frac{\pi}{2N-1}} & = \left ( \frac{2K}{\pi^2} -\frac{M}{2}\right )N.
\end{split}
\end{equation}
Therefore,
\begin{equation}
\lim\limits_{N \rightarrow\infty}\frac{K \left (\lambda_0-\frac{|\lambda_1|}{2\cos^2 \frac{\pi}{2N-1}} \right )}{N\cdot \left[ (2N-1)K-1-\frac{1}{2\cos^2 \frac{\pi}{2N-1}}\right ]}=\frac{3M}{4K}+\frac{1}{2}-\frac{1}{\pi^2}.
\end{equation}
On the other hand, let us rewrite the Welch bound expression (\ref{Welch_bound_for_cc})  as
\begin{equation}\label{decomposed-welch-bound}
M^2N^2\frac{\frac{K}{M}-1}{K(2N-1)-1} = M \left ( N- \mathcal{R}_1 \right )
\end{equation}
with
\begin{equation}\label{R1}
\mathcal{R}_1 \triangleq \frac{N(MN-1)+N(N-1)K}{(2N-1)K-1}.
\end{equation}
Then,
\begin{equation}\label{ch4_R1_largeN}
\lim\limits_{N\rightarrow \infty}\frac{\mathcal{R}_1}{N}=\frac{1}{2}+\frac{M}{2K}.
\end{equation}
With (\ref{R1}) and (\ref{ch4_R1_largeN}), one can show that the lower bound in \textit{Lemma \ref{coro_GLB_from_wetvec3}} is asymptotically tighter than the Welch bound in (\ref{Welch_bound_for_cc}) if and only if the following equation is satisfied.
\begin{equation}\label{asym_bd_wtvec3}
\frac{1}{2}+\frac{M}{2K}>\frac{3M}{4K}+\frac{1}{2}-\frac{1}{\pi^2}.
\end{equation}
Equivalently, we need to prove that for $K=\left \lfloor \frac{\pi^2M}{4}\right \rfloor+1$ (as $N \rightarrow\infty$), the following inequality holds.
\begin{equation}\label{BdfromOptWtVec2}
d_1(M)\triangleq \frac{\left \lfloor \frac{\pi^2M}{4}\right \rfloor+1}{M}- \frac{\pi^2}{4} > 0.
\end{equation}
One can readily show that the condition $d_1(M)>0$ given in (\ref{BdfromOptWtVec2}) is true for \textit{all} $M\geq 2$. Therefore, we have the following theorem.

\vspace{0.1in}

\begin{theorem}\label{th4OptWtVec}
The GLB in (\ref{coro_GLB_from_wetvec3-equ}) which arises from the weight vector in (\ref{wgtvec3_equ}) reduces to
\begin{equation}\label{aym_lwrbd_wec3_equ}
\delta^2_{\max} \gtrsim MN\left [ \left ( \frac{1}{2}+\frac{1}{\pi^2}\right ) -\frac{3M}{4K}\right ],
\end{equation}
for sufficiently large $N$. Such an asymptotic lower bound is tighter than the Welch bound for \textit{all} $K\geq \overline{K}+1$ and for \textit{all} $M\geq 2$. 
\end{theorem}

\subsection{Proof of Local Optimality}
In this subsection, we prove the proposed weight vector in (\ref{wgtvec3_equ}) is a local minimizer of the GLB in (\ref{generalized_welch_bound_for_cc-equ}) under certain condition. We consider the weight vector $\mathbf{w}$  by setting $j=0$ in (\ref{wgtvec3_equ}) because other values of $j$ will lead to identical value of GLB [cf. (\ref{GLB_quadra_fd}) and (\ref{GLB_quadra_fd_})].

\begin{equation}\label{proposed_wgtvec}
w_i=\frac{1}{2N-1}\left ( 1+\frac{\cos \frac{2\pi i}{2N-1}}{\cos \frac{\pi}{2N-1}}\right ),~i\in\{0,1,\cdots,2N-2\}.
\end{equation}
Note that the frequency domain vector $\mathbf{v}=\mathbf{F}_{2N-1}\mathbf{w}$ has $v_0=1, v_1=v_{2N-2}=\frac{1}{2\cos \frac{\pi}{2N-1}}$ and $v_l=0$ for all $l\in\{2,3,,\cdots,2N-3\}$. Our problem in this subsection can be formally cast as follows.

\vspace{0.1in}

\begin{problem}
Define the fractional quadratic function $f(\mathbf{x})$\footnote{Note that $f(\mathbf{x})$ is essentially the fractional quadratic term in (\ref{generalized_welch_bound_for_cc-equ}) by replacing $\mathbf{w}$ with $\mathbf{x}$.} as follows.
\begin{equation}\label{FQP_defi}
f(\mathbf{x})\triangleq\frac{\mathbf{x}^{\text{T}}\mathbf{Q}_a\mathbf{x}}{1-\frac{1}{K}\cdot \mathbf{x}^{\text{T}}\mathbf{x}},\\
\end{equation}
where $x_i\geq0, i\in\{0,1,\cdots,2N-2\},\sum _{i=0}^{2N-2}x_i=1$, $\mathbf{Q}_a$ is the circulant matrix defined in (\ref{Leven-quad-fun-equ}) which has order $(2N-1)$ and with $a=(MN-1)N/K$. When $K=\overline{K}+1$ and $M,N$ becomes sufficiently large, prove that the weight vector $\mathbf{w}$ in (\ref{proposed_wgtvec}) is a local minimizer of $f(\mathbf{x})$, i.e.,
\begin{equation}\label{local_mini_equ}
f(\mathbf{w}+\mathbf{e})\geq f(\mathbf{w}),
\end{equation}
 holds for any feasible perturbation $\mathbf{e}$ which has sufficiently small norm.
\end{problem}

\begin{proof}
To get started, we define
\begin{equation}
\begin{split}
\alpha(\mathbf{w},\mathbf{e}) & \triangleq \mathbf{w}^{\text{T}}\mathbf{Q}_a\mathbf{w}\mathbf{e}^{\text{T}}\mathbf{e}-\mathbf{w}^{\text{T}}\mathbf{w}\mathbf{e}^{\text{T}}\mathbf{Q}_a\mathbf{e},\\
\beta(\mathbf{w},\mathbf{e}) & \triangleq \mathbf{w}^{\text{T}}\mathbf{Q}_a\mathbf{w}\mathbf{e}^{\text{T}}\mathbf{w}-\mathbf{w}^{\text{T}}\mathbf{w}\mathbf{w}^{\text{T}}\mathbf{Q}_a\mathbf{e},\\
\gamma(\mathbf{w},\mathbf{e}) & \triangleq \frac{\alpha(\mathbf{w},\mathbf{e})+2\beta(\mathbf{w},\mathbf{e})}{K}.
\end{split}
\end{equation}
It is easy to show that (\ref{local_mini_equ}) is equivalent to the following inequality.
\begin{equation}\label{local_mini_equ2}
2\mathbf{w}^{\text{T}}\mathbf{Q}_a\mathbf{e}+\mathbf{e}^{\text{T}}\mathbf{Q}_a\mathbf{e}+\gamma(\mathbf{w},\mathbf{e})\geq 0.
\end{equation}
Let $\mathbf{E}=\mathbf{F}_{2N-1}\mathbf{e}$. Since $\mathbf{e}$ is a real vector, $\mathbf{E}$ is conjugate symmetric in that $E_l=E^*_{2N-1-l}$ for $l=1,2,\cdots,2N-2$. By taking advantage of (\ref{Qa_in_FreqDomain}), we present the following properties which will be useful in the sequel.
\begin{subequations}
\begin{align}
 & E_0=\sum\limits_{i=0}^{2N-2}e_i=0;\label{multi_equ1}\\
 & \mathbf{w}+\mathbf{e}\geq 0;\label{multi_equ2}\\
 & \mathbf{w}^{\text{T}}\mathbf{Q}_a\mathbf{e}=\lambda_1 \cdot \frac{E_1+E^*_1}{2(2N-1)\cos \frac{\pi}{2N-1}};\label{multi_equ3}\\
 & \mathbf{e}^{\text{T}}\mathbf{Q}_a\mathbf{e}=\frac{2}{2N-1}\cdot \sum\limits_{i=1}^{N-1}\lambda_i |E_i|^2; \label{multi_equ4}\\
 & \mathbf{w}^{\text{T}}\mathbf{Q}_a\mathbf{w}=\frac{1}{2N-1}\left ( \lambda_0+\frac{\lambda_1}{2\cos^2\frac{\pi}{2N-1}}\right );\label{multi_equ5}\\
 & \mathbf{e}^{\text{T}}\mathbf{w}=\frac{E_1+E^*_1}{2(2N-1)\cos \frac{\pi}{2N-1}};\label{multi_equ6}\\
 & \mathbf{w}^{\text{T}}\mathbf{w}=\frac{1}{2N-1}\left ( 1+\frac{1}{2\cos^2\frac{\pi}{2N-1}}\right );\label{multi_equ7}\\
 & \mathbf{e}^{\text{T}}\mathbf{e}=\frac{2}{2N-1}\sum\limits_{i=1}^{N-1}|E_i|^2\label{multi_equ8}.
\end{align}
\end{subequations}
By (\ref{multi_equ4}), (\ref{multi_equ5}), (\ref{multi_equ7}) and (\ref{multi_equ8}), we have
\begin{equation}
\begin{split}
\alpha(\mathbf{w},\mathbf{e})& =\frac{2}{(2N-1)^2}\cdot \Bigl \{ (\lambda_0-\lambda_1)|E_1|^2  \\
                            & \left. + \sum\limits_{i=2}^{N-1} \left [ (\lambda_0-\lambda_1) + \frac{\lambda_1-\lambda_i}{2\cos^2 \frac{\pi}{2N-1}} \right ] |E_i|^2\right \}.
\end{split}
\end{equation}
By (\ref{multi_equ3}), (\ref{multi_equ5}), (\ref{multi_equ6}) and (\ref{multi_equ7}), we have
\begin{equation}
\beta(\mathbf{w},\mathbf{e})=\frac{\lambda_0-\lambda_1}{2(2N-1)^2}\cdot \frac{E_1+E^*_1}{\cos \frac{\pi}{2N-1}}.
\end{equation}
Therefore, $\gamma(\mathbf{w},\mathbf{e})$ can be expressed in the form shown in (\ref{gamma_equ}).
\begin{figure*}
\begin{equation}\label{gamma_equ}
\begin{split}
\gamma(\mathbf{w},\mathbf{e}) &  =  \frac{\lambda_1}{K(2N-1)^2} \cdot \left \{ \sum\limits_{i=2}^{N-1}\frac{|E_i|^2}{\cos^2 \frac{\pi}{2N-1}}-2|E_1|^2-\frac{E_1+E^*_1}{\cos \frac{\pi}{2N-1}}\right \}\\
 & ~+\frac{1}{K(2N-1)^2} \cdot \left \{ 2\lambda_0 |E_1|^2+ \sum\limits_{i=2}^{N-1}\left ( \lambda_0-\lambda_i-\frac{\lambda_i}{\cos^2\frac{\pi}{2N-1}}\right ) |E_i|^2+\lambda_0 \frac{E_1+E^*_1}{\cos \frac{\pi}{2N-1}}\right \}.
\end{split}
\end{equation}
\end{figure*}
Since $\mathbf{e}$ is a small perturbation, let us assume
\begin{equation}\label{rho_defi_equ}
0\leq 2\sum\limits_{i=1}^{N-1}|E_i|^2\ll 1.
\end{equation}
Next, we proceed with the following two cases.
\begin{enumerate}
\item Case I: If there exists $E_i\neq 0$ for $i\in\{2,3,\cdots,N-1\}$.

Since we consider $K=\overline{K}+1$ with sufficiently large $M,N$, it is readily to show that $\lambda_i>0$ holds for any $i\in\{2,3,\cdots,N-1\}$ [see (\ref{lambda_even_equ}) and (\ref{lambda_odd_equ})]. By (\ref{multi_equ4}), let us write
\begin{equation}
\mathbf{e}^{\text{T}}\mathbf{Q}_a\mathbf{e}(2N-1)=2\lambda_1 |E_1|^2+\xi,
\end{equation}
where $\xi=2\sum\limits_{i=2}^{N-1}\lambda_i |E_i|^2>0$. Furthermore, write
\begin{equation}\label{local_mini_equ3}
\Bigl[ 2\mathbf{w}^{\text{T}}\mathbf{Q}_a\mathbf{e}+\mathbf{e}^{\text{T}}\mathbf{Q}_a\mathbf{e}+\gamma(\mathbf{w},\mathbf{e}) \Bigl ]\cdot (2N-1)=\lambda_1 A+B,
\end{equation}
where
\begin{equation}\label{AB_equ}
\begin{split}
A  = & 2\left ( 1-\frac{1}{K(2N-1)}\right )|E_1|^2\\
     & ~~~+ \left ( 1-\frac{1}{K(2N-1)}\right )\cdot\frac{E_1+E^*_1}{\cos \frac{\pi}{2N-1}}\\
     & ~~~+\frac{1}{K(2N-1)} \sum\limits_{i=2}^{N-1}\frac{|E_i|^2}{\cos^2 \frac{\pi}{2N-1}},\\
B = & \xi + \frac{1}{K(2N-1)} \cdot  2\lambda_0 |E_1|^2\\
    & +\frac{1}{K(2N-1)} \cdot \sum\limits_{i=2}^{N-1}\left ( \lambda_0-\lambda_i-\frac{\lambda_i}{\cos^2\frac{\pi}{2N-1}}\right ) |E_i|^2\\
    &+\frac{1}{K(2N-1)} \cdot \lambda_0 \frac{E_1+E^*_1}{\cos \frac{\pi}{2N-1}}.
\end{split}
\end{equation}
\vspace{0.1in}
\begin{remark}\label{rmk_on_AB}
Since $K=\left \lfloor 4(MN-1)N\sin^2 \frac{\pi}{2(2N-1)} \right \rfloor+1$, $A$ and $B$ approach to \\$\left (2|E_1|^2+E_1+E^*_1 \right)$ and $\xi$, respectively, as $M$ grows sufficiently large.
\end{remark}
\vspace{0.1in}

To show (\ref{local_mini_equ}) [and (\ref{local_mini_equ2})] holds, we only need to prove the right-hand term of (\ref{local_mini_equ3}) divided by $a$ is nonnegative, asymptotically. For this, our idea is to consider a fixed $N$ (sufficiently large) and prove that: (1) $\lim\limits_{M\rightarrow + \infty} \frac{B}{a}$ is lower bounded by a nonnegative value determined by $N$ only; (2) $\lim\limits_{M\rightarrow + \infty}\frac{\lambda_1}{a}$ tends to zero (with an upper bounded $\lim\limits_{M\rightarrow + \infty}A$) regardless the value of $N$.

From (\ref{lambda_i_equ}), we have
\begin{equation}
\frac{\lambda_{2i}}{a}=1-\frac{1}{4a\cos^2 \frac{\pi(2i)}{2(2N-1)}},~2\leq 2i \leq N-1.
\end{equation}
For ease of analysis, let $N$ be an even integer\footnote{When $N$ is odd, we can prove (\ref{local_mini_equ}) [and (\ref{local_mini_equ2})] holds by almost the same arguments.}. Hence, $\max\limits_{2\leq 2i \leq N-1}(2i)=N-2$. Since $\frac{\lambda_{2i}}{a}$ is a decreasing function of $i$, we have
\begin{equation}\label{lambda_even_equ}
\frac{\lambda_{2i}}{a}\geq 1-\frac{1}{4a\cos^2 \frac{\pi(N-2)}{2(2N-1)}}>2/3.
\end{equation}
Also,
\begin{equation}
\frac{\lambda_{2i+1}}{a}=1-\frac{1}{4a\sin^2 \frac{\pi(2i+1)}{2(2N-1)}}\geq 1- \frac{1}{4a \sin^2 \frac{3\pi}{2(2N-1)}},
\end{equation}
where $2\leq 2i+1\leq N-1.$ By noting $\sin 3x > 2\sin x$ ($x$ a small positive angle) and $K=\overline{K}+1$, we have
\begin{equation}\label{lambda_odd_equ}
\begin{split}
\frac{\lambda_{2i+1}}{a}  & > 1-\frac{1}{4}\cdot \frac{\left \lfloor 4(MN-1)N\sin^2 \frac{\pi}{2(2N-1)} \right \rfloor+1}{4(MN-1)N\sin^2 \frac{\pi}{2(2N-1)}}\\
& > 1-\frac{1}{4}\cdot\frac{4}{3}=\frac{2}{3}.
\end{split}
\end{equation}
By (\ref{lambda_even_equ}) and (\ref{lambda_odd_equ}), we obtain
\begin{equation}\label{lim_xi_a_equ}
\lim\limits_{M\rightarrow + \infty} \frac{B}{a}=\frac{\xi}{a}=\sum\limits_{i=2}^{N-1}\left (\frac{\lambda_i}{a} \right ) \cdot 2|E_i|^2 \geq \frac{2}{3}\cdot \left ( 2\sum\limits_{i=2}^{N-1}|E_i|^2 \right ).
\end{equation}
On the other hand,
\begin{equation}\label{lambda1_equ}
\begin{split}
 & \lim\limits_{M\rightarrow+\infty}\frac{\lambda_1}{a} \\
 = & \lim\limits_{M\rightarrow+\infty}\left [1-\frac{\left \lfloor 4(MN-1)N\sin^2 \frac{\pi}{2(2N-1)} \right \rfloor+1}{4N(MN-1)\sin^2 \frac{\pi}{2(2N-1)}} \right ]\\
 \rightarrow & ~ 0^{-},
\end{split}
\end{equation}
where $0^{-}$ denotes a sufficiently small value (negative) that approaches zero from the left.
Therefore, we have
\begin{equation}\label{lim_gep_equ}
\begin{split}
  & \lim\limits_{M\rightarrow +\infty}\frac{\Bigl[ 2\mathbf{w}^{\text{T}}\mathbf{Q}_a\mathbf{e}+\mathbf{e}^{\text{T}}\mathbf{Q}_a\mathbf{e}+\gamma(\mathbf{w},\mathbf{e}) \Bigl ]\cdot (2N-1)}{a}\\
  = & {\lim\limits_{M\rightarrow+\infty}\frac{\lambda_1}{a}} \cdot \underbrace{\lim\limits_{M\rightarrow+\infty}A}_{\text{upper bounded}} + {\lim\limits_{M\rightarrow+\infty}\frac{B}{a}}
\end{split}
\end{equation}
By (\ref{lim_xi_a_equ}) and (\ref{lambda1_equ}), we assert that when $M$ is sufficiently large, the sign of the limit in (\ref{lim_gep_equ}) will be identical to that of $\xi/a$ [cf. (\ref{lim_xi_a_equ})] which is nonnegative. This shows that (\ref{local_mini_equ}) [and (\ref{local_mini_equ2})] holds for Case I, asymptotically.

\vspace{0.1in}

\item Case II: If $E_i=0$ for all $i\in\{2,3,\cdots,N-1\}$.

In this case, (\ref{AB_equ}) reduces to
\begin{equation}\label{AB_equ2}
\begin{split}
A  = & 2\left ( 1-\frac{1}{K(2N-1)}\right )|E_1|^2\\
     & + \left ( 1-\frac{1}{K(2N-1)}\right )\frac{E_1+E^*_1}{\cos \frac{\pi}{2N-1}},\\
B = & \frac{\lambda_0}{K(2N-1)} \cdot \left \{ 2 |E_1|^2+ \frac{E_1+E^*_1}{\cos \frac{\pi}{2N-1}}\right \}.
\end{split}
\end{equation}
Since $\mathbf{E}=\mathbf{F}_{2N-1}\mathbf{e}$, we have
\begin{equation}
e_i=\frac{2}{2N-1}\text{Re}\left \{E_1\exp\left ( \frac{\sqrt{-1}2\pi i }{2N-1}\right ) \right \},
\end{equation}
where $\text{Re}\{x\}$ denotes the real part of complex data $x$.
Consider $E_1$ which takes the following form.
\begin{equation}\label{Ei_equ}
E_1=\frac{t}{2\cos \frac{\pi}{2N-1}}\exp\left ( \sqrt{-1}\psi\right ),
\end{equation}
where $0\leq t\ll 1$ and $\psi$ denotes the phase shift of $E_1$. As a result, $e_i$ can be expressed as
\begin{equation}
e_i=\frac{t}{(2N-1)\cos \frac{\pi}{2N-1}} \cdot \cos\left ( \frac{2\pi i}{2N-1}+\psi\right ).
\end{equation}
Thus,
\begin{equation}\label{lambda1AB_CaseII}
\begin{split}
 & \lambda_1 A+B\\
 = & \left ( 2 |E_1|^2+ \frac{E_1+E^*_1}{\cos \frac{\pi}{2N-1}} \right ) \cdot \left ( \lambda_1 + \frac{\lambda_0-\lambda_1}{K(2N-1)}\right ).
\end{split}
\end{equation}
Since $\lambda_1\thicksim  O(N^2),\frac{\lambda_0-\lambda_1}{K(2N-1)}\thicksim O(\frac{N}{K})\thicksim O(\frac{N}{M})$, we assert that for sufficiently large $M,N$,
\begin{equation}\label{lambda1AB_CaseII_2}
\lambda_1+\frac{\lambda_0-\lambda_1}{K(2N-1)}<0,
\end{equation}
holds because it will be dominated by the negative $\lambda_1$. Our next task is to show that $\left ( 2 |E_1|^2+ \frac{E_1+E^*_1}{\cos \frac{\pi}{2N-1}} \right )\leq 0$. By (\ref{Ei_equ}), we have
\begin{equation}\label{lambda1AB_CaseII_3}
2 |E_1|^2+ \frac{E_1+E^*_1}{\cos \frac{\pi}{2N-1}} =\frac{1}{2\cos^2\frac{\pi}{2N-1}}(t^2+2t\cos \psi).
\end{equation}
It is required in (\ref{multi_equ2}) that $w_i+e_i\geq 0$ for all $i$, i.e.,
\begin{equation}
\cos \frac{\pi}{2N-1} + \cos \frac{2 \pi i}{2N-1} + t \cos \left ( \frac{2\pi i}{2N-1} +\psi\right )\geq 0.
\end{equation}
Setting $i=N$, we have
\begin{equation}
\begin{split}
 & \cos \left ( \frac{2\pi N}{2N-1} +\psi\right )\geq 0 \\
 \rightarrow & \left ( \frac{1}{2}-\frac{1}{2N-1}\right )\pi \leq \psi \leq \left ( \frac{3}{2}-\frac{1}{2N-1}\right ).
\end{split}
\end{equation}
Setting $i=N$, we have
\begin{equation}
\begin{split}
 & \cos \left ( \frac{2\pi (N-1)}{2N-1} +\psi\right )\geq 0 \\
 \rightarrow & \left ( \frac{1}{2}+\frac{1}{2N-1}\right )\pi \leq \psi \leq \left ( \frac{3}{2}+\frac{1}{2N-1}\right ).
\end{split}
\end{equation}
Therefore,
\begin{equation}
\begin{split}
 & \left ( \frac{1}{2}+\frac{1}{2N-1}\right )\pi \leq \psi \leq \left ( \frac{3}{2}-\frac{1}{2N-1}\right ) \\
 \rightarrow & -1\leq \cos \psi <0.
\end{split}
\end{equation}
This shows $t^2+2t \cos \psi\leq 0$ holds provided $t\leq -2\cos \psi$. This can be easily satisfied by a sufficiently small $t$. Together with (\ref{lambda1AB_CaseII})-(\ref{lambda1AB_CaseII_3}), we conclude that (\ref{local_mini_equ}) [and (\ref{local_mini_equ2})] holds for Case II, asymptotically. This completes the proof of the local optimality of the proposed weight vector in (\ref{proposed_wgtvec}).
\end{enumerate}
\end{proof}

\begin{remark}
Following a proof similar to the above, one can easily show that the weight vector $\mathbf{w}$ in (\ref{proposed_wgtvec}) is also a local minimizer of the constrained QP of $\min\limits_{\mathbf{w}} {Q}\left(\mathbf{{w}},\frac{N(MN-1)}{K}\right)$ when $K=\overline{K}+1$ and $M,N$ are sufficiently large.
\end{remark}

\section{Discussions and Comparisons}

In this section, we first consider another two weight vectors and study the tightness of their resultant GLBs. Then, we compare them with the proposed weight vector in (\ref{wgtvec3_equ}) by some numerical results.

\subsection{GLB from Weight Vector 2}

In \cite{LiuParaGuanBozas14}, Liu \emph{et al} showed that the following ``positive-cycle-of-sine" weight vector $\mathbf{w}$
\begin{equation}\label{sine_shape_weight_vector}
{w}_i=\left \{
\begin{array}{ll}
\tan \frac{\pi}{2m} \sin \frac{\pi i}{m}, & ~~i\in \{0,1,\cdots,m-1\};\\
0,   & ~~i\in \{m,m+1,\cdots,2N-2\},
\end{array}
\right .
\end{equation}
where $2\leq m \leq 2N-1$, asymptotically leads to a tighter Levenshtein bound (i.e., $M=1$) for all $K\geq3$ \cite{Levenshtein99}. 

\vspace{0.1in}

By [\ref{LiuParaGuanBozas14}, \textit{Proposition 1}], one can show that the resultant GLB from the weight vector in (\ref{sine_shape_weight_vector}) can be written as follows.

\begin{corollary}\label{new_LwerBd_from_new_wv}
\begin{equation}\label{ZL-corollary-new-weight-vector-equ}
\delta^2_{\max}\geq   M \left [
N-\frac{N(MN-1)m\tan^2\frac{\pi}{2m}+2KQ({{\mathbf{{w}}}},0)}{2K-m\tan^2\frac{\pi}{2m}}
\right ], 
\end{equation}
where 
\begin{equation}
\begin{split}
  & Q(\mathbf{w},0)   \\
= &
\left\{
\begin{array}{ll}
\frac{m}{4}\left(1-\tan^2\frac{\pi}{2m} \right ), \\
~~~~~~~~~~~~~~~~~~~~~~~~~~~~~\text{for}~2 \leq m \leq N, \\
-\frac{3m-4N+2}{4} -\frac{m}{4}\tan^2\frac{\pi}{2m}+\frac{ m-N-1}{2} \cos \frac{N\pi}{m}\\
+ \left ( \frac{2m-2N+1}{4}\tan\frac{\pi}{2m}+\frac{3}{4\tan\frac{\pi}{2m}} \right )\sin\frac{N\pi}{m}, \\
~~~~~~~~~~~~~~~~~~~~~~~~~~~~~\text{for}~N<m\leq 2N-1.
\end{array}
\right.
\end{split}
\end{equation}
\end{corollary}

\vspace{0.1in}

In what follows, we analyze the asymptotic tightness of the lower bound in (\ref{ZL-corollary-new-weight-vector-equ}).

Define $r \triangleq \lim\limits_{N\rightarrow \infty}
 m/N$. Obviously, $r$ is a real-valued constant with $0<r<2$ when $m$ is on the same order of $rN$ (i.e., $m\sim rN$);  and $r\rightarrow 0$ when $m$ is dominated by $N$ asymptotically (i.e., $m \sim o(N)$). Furthermore, define the fractional term in (\ref{ZL-corollary-new-weight-vector-equ}) as
\begin{equation}\label{R2}
\mathcal{R}_2 \triangleq \frac{N(MN-1)m\tan^2\frac{\pi}{2m}+2KQ({\mathbf{{w}}},0)}{2K-m\tan^2\frac{\pi}{2m}}.
\end{equation}
It is easy to see that the lower bound in (\ref{ZL-corollary-new-weight-vector-equ}) is tighter than the Welch
bound in (\ref{Welch_bound_for_cc}) if and only if
\begin{equation}\label{iff}
\mathcal{R}_1>\min\limits_{2\leq m \leq 2N-1} \mathcal{R}_2,
\end{equation}
where $\mathcal{R}_1$ is defined in (\ref{R1}). As $N$ tends to infinity, the inequality in (\ref{iff}) is equivalent to
\begin{equation}\label{asymp_iff}
\lim\limits_{N\rightarrow \infty}\frac{\mathcal{R}_1}{N} >
\lim\limits_{N\rightarrow \infty} \min\limits_{2\leq m \leq 2N-1} \frac{\mathcal{R}_2}{N}.
\end{equation}

When $m \sim o(N)$, we have $r\rightarrow 0$ and $rN \in [2,\infty)$ as $N\rightarrow\infty$. In this case, one can show that 
\begin{equation}
\begin{split}
 &\lim\limits_{N \rightarrow\infty} \frac{\mathcal{R}_2}{N} \\
 =&\lim\limits_{N \rightarrow\infty}  \frac{N(MN-1)rN \tan^2\frac{\pi}{2rN}+2K\cdot\frac{rN}{4}\left(1-\tan^2\frac{\pi}{2rN} \right )}{N(2K-rN\tan^2\frac{\pi}{2rN})}  \\
 =& \lim\limits_{N \rightarrow\infty}  \frac{MN(rN \tan^2\frac{\pi}{2rN})+\frac{Kr}{2}\left(1-\tan^2\frac{\pi}{2rN} \right )}{2K-rN\tan^2\frac{\pi}{2rN}} \\
 =& \left\{
\begin{array}{ll}
\infty, & \text{for}~2 \leq rN < \infty; \\
\frac{M\pi^2}{8Kr}+\frac{r}{4} \rightarrow \infty,  & \text{for}~rN \rightarrow \infty,
\end{array}
\right.
\end{split}
\end{equation}
which can be ignored without missing the minimum point of interest in the right-hand side of (\ref{asymp_iff}).
Hence, we shall assume $r$ to be a non-vanishing real-valued constant with $0 < r< 2$, and  rewrite (\ref{asymp_iff}) as
\begin{equation}\label{asymp_iff_}
\lim\limits_{N\rightarrow \infty}\frac{\mathcal{R}_1}{N} >
\min\limits_{0<r<2} \lim\limits_{N\rightarrow \infty}  \frac{\mathcal{R}_2}{N}.
\end{equation}
Here, the order of the limit and minimization operations can be exchanged because $\lim\limits_{N\rightarrow \infty}\frac{\mathcal{R}_2}{N}$ as a function of $r$ exists,  as shown below.
Next, noting that $\lim\limits_{N\rightarrow \infty}m\tan^2 \frac{\pi}{2m}=\lim\limits_{N\rightarrow \infty}rN\tan^2\frac{\pi}{2rN}=0$, we can express (\ref{R2}) as
\begin{equation}\label{R2_2}
\begin{split}
\lim\limits_{N \rightarrow\infty} \frac{\mathcal{R}_2}{N} & =\lim\limits_{N \rightarrow\infty}\frac{MN-1}{2K}m\tan^2\frac{\pi}{2m}+\lim\limits_{N \rightarrow\infty}\frac{Q({{\textbf{{w}}}},0)}{N},
\end{split}
\end{equation}
where
\begin{equation}\label{R2_2a}
\begin{split}
\lim\limits_{N \rightarrow\infty}\frac{MN-1}{2K}m\tan^2\frac{\pi}{2m}=\frac{M\pi^2}{8Kr},
\end{split}
\end{equation}
and after some manipulations,
\begin{equation}\label{ch4_Q_largeN}
\begin{split}
f(r) & \triangleq \lim\limits_{N \rightarrow\infty}\frac{Q({{\textbf{{w}}}},0)}{N}\\
   & = \left \{
\begin{array}{cl}
r/4, & ~~\text{for}~0< r\leq 1; \\
\frac{4-3r}{4} + \frac{r-1}{2}\cos\frac{\pi}{r} + \frac{3r}{2\pi}\sin\frac{\pi}{r},  &~~\text{for}~1<r< 2.
\end{array}
\right.
\end{split}
\end{equation}
By (\ref{ch4_R1_largeN}), (\ref{R2_2a}) and (\ref{ch4_Q_largeN}), it follows that (\ref{asymp_iff_}) reduces to
\begin{equation}\label{asym_bd_wtvec2}
\frac{1}{2} + \frac{M}{2K} > \min_{0<r<2} \left(\frac{M\pi^2}{8Kr} + f(r)\right).
\end{equation}
Equivalently, we assert that the asymptotic lower bound in (\ref{ZL-corollary-new-weight-vector-equ}) is tighter than the Welch
bound if and only if
\begin{equation}\label{asym_bd_wtvec2_equ}
 \frac{K}{M} > \min_{0< r < 2} L(r)\approx 2.483257,
\end{equation}
where
\begin{equation}
{L}(r) \triangleq
\left \{
\begin{array}{ll}
\frac{\pi^2-4r}{4r-2r^2}, & \text{for}~0<r\leq 1;\\
\frac{\pi^2-4r}{2r(3r-2)-4r(r-1)\cos\frac{\pi}{r}-12\frac{r^2}{\pi}\sin \frac{\pi}{r}},  & \text{for}~1<r< 2.
\end{array}
\right.
\end{equation}

\begin{figure}[!ht]
\centering
\scalebox{0.55}
{\includegraphics{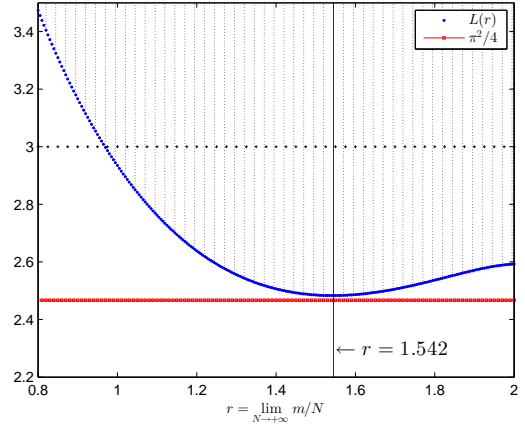}}
\caption{A plot of $L(r)$ and $\pi^2/4$ versus $r$.}
\label{Fig_optimal_r2}
\end{figure}

In Fig. \ref{Fig_optimal_r2}, $L(r)$ and $\pi^2/4$ versus $r$
over the range of $0.8\leq r \leq 2$ are plotted.
It can be obtained from (\ref{overlineK}) and Fig. \ref{Fig_optimal_r2} that
\begin{equation}
\frac{\lim\limits_{N\rightarrow\infty}\overline{K}}{M}
\leq \underbrace{\frac{\pi^2}{4}}_{\approx 2.467401} <  ~\underbrace{\min_{0< r < 2} {L}(r)}_{\approx 2.483257}.
\end{equation}

By (\ref{asym_bd_wtvec2_equ}), one can see that the proposed weight vector in (\ref{sine_shape_weight_vector}) asymptotically leads to a tighter GLB for \textit{all} $K\geq \overline{K}+1$ if and only if the value of $M$ satisfies the following condition [c.f.~(\ref{nece_cond_QCSSBd2_2})]
\begin{equation}\label{ch5_dM_equ}
d_2(M) \triangleq \frac{\left \lfloor \frac{\pi^2M}{4}\right \rfloor+1}{M}- \min_{0< r < 2} {L}(r) > 0.
\end{equation}

\vspace{0.1in}

\begin{figure*}[!ht]
\centerline{
\subfloat[$2\leq M \leq 512$]{\includegraphics[width=3in]{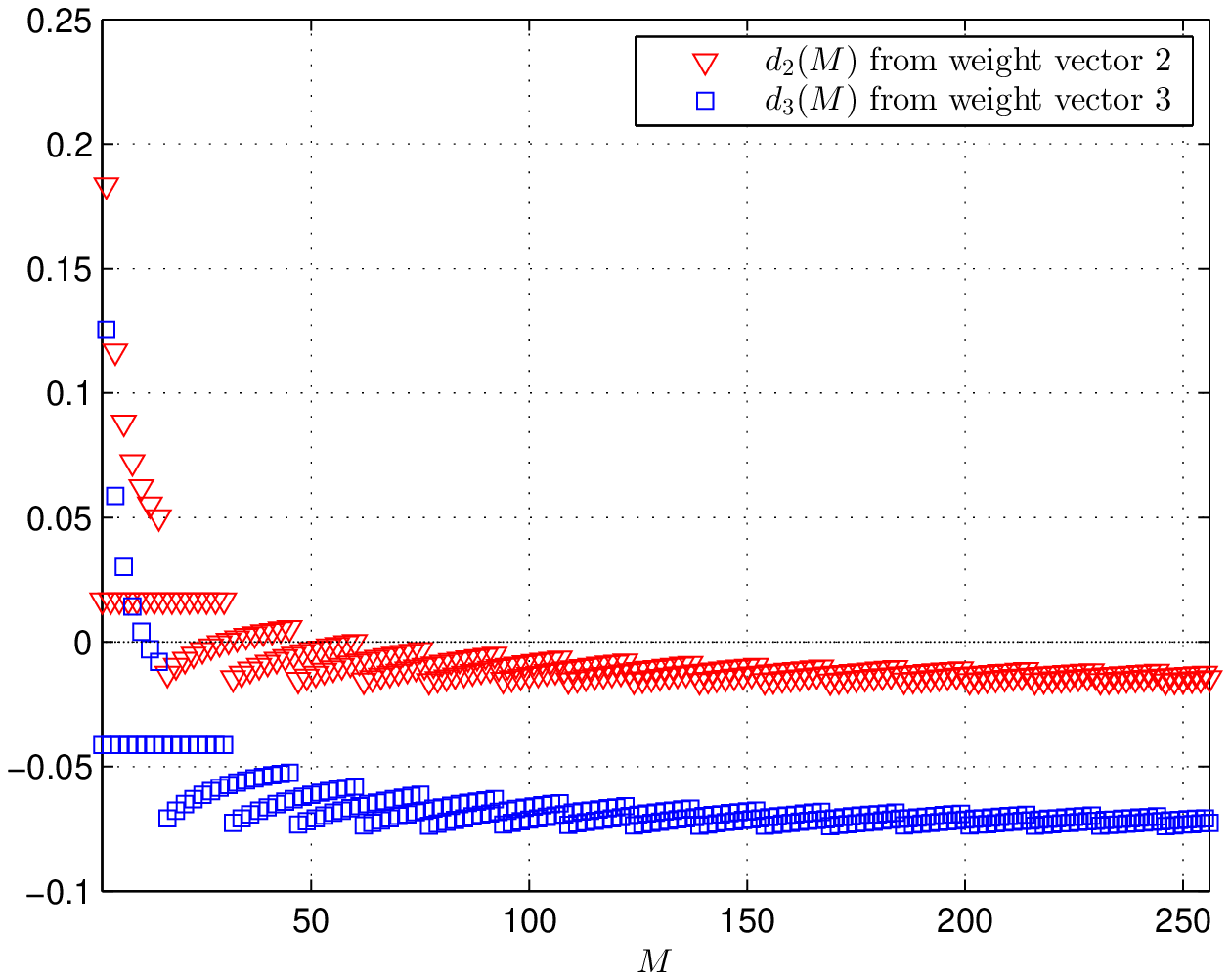}
\label{fig_second_case}}
\hfil
\subfloat[$2\leq M \leq 80$]{\includegraphics[width=3in]{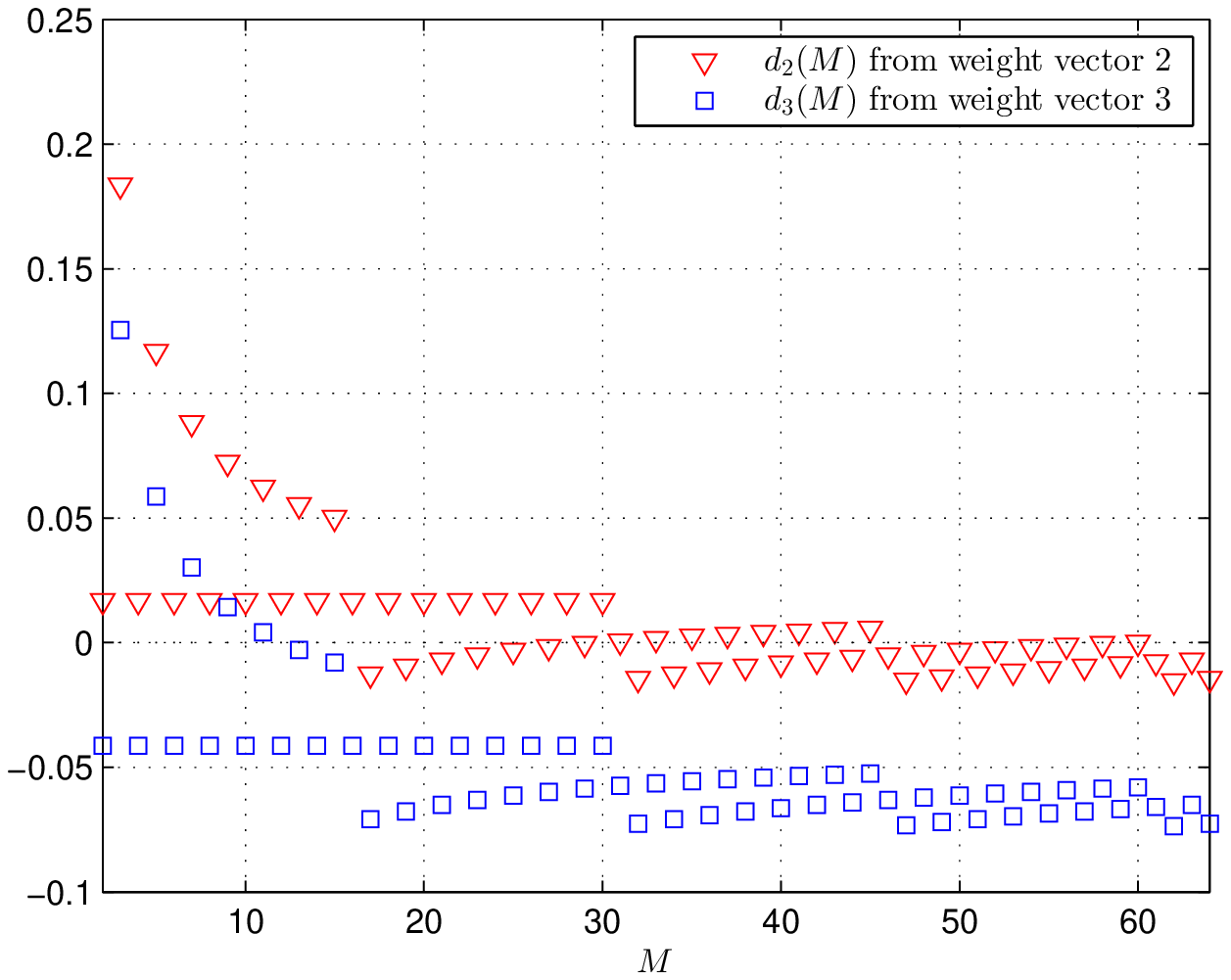}
\label{fig_first_case}}
}
\caption{A plot of $d_2(M)$ in (\ref{ch5_dM_equ}) and $d_3(M)$ in (\ref{ch5_dM_equ2}) versus $M$, where subplot (a) is a zoom-in of subplot (b). It is noted that a positive $d_2(M)$ [or $d_3(M)$] corresponds to a tighter GLB over the Welch bound.}
\label{fig_dM}
\end{figure*}

In Fig. \ref{fig_dM}, $d_2(M)$ versus $M$ is also plotted. By identifying $M$ satisfying $d_2(M)>0$ [shown in (\ref{ch5_dM_equ})],  we arrive at the following theorem.

\vspace{0.1in}
\begin{theorem}\label{tighter_M_rmk}
The GLB in (\ref{ZL-corollary-new-weight-vector-equ}) which arises from the weight vector in (\ref{sine_shape_weight_vector}) reduces to
\begin{equation}\label{aym_lwrbd_wec2_equ}
\delta^2_{\max}\gtrsim   MN \left [1-\min_{0<r<2} \left(\frac{M\pi^2}{8Kr} + f(r)\right)\right ],
\end{equation}
for sufficiently large $N$, where $f(r)$ is given in (\ref{ch4_Q_largeN}). Such an asymptotic lower bound is tighter than the Welch bound for \textit{all} $K\geq \overline{K}+1$ if and only if
\begin{equation}\label{distri_optimal_M}
\begin{split}
 M \in & \Bigl \{2,3,4,5,6,7,8,9,10,11,12,13,14,15,16,18,20,\\
 & ~~~22,24, 26,28,30,31,33, 35,37,39,41,43,45,60 \Bigl \}.
\end{split}
\end{equation}
\end{theorem}

\vspace{0.1in}

\subsection{GLB from Weight Vector 3}

Let us consider the weight vector obtained by minimizing the following function using the Lagrange multiplier.
\begin{equation}\label{mini_F}
\mathcal{F}_{K,M,N,m}(\mathbf{w})=Q\left(\mathbf{w},\frac{MN^2}{K}\right)-2\lambda\left (\sum\limits_{i=0}^{m-1}w_i-1 \right),
\end{equation}
where $w_i=0$ for $i\in \{m,m+1,\cdots,2N-2\}$ and $2\leq m \leq 2N-1$. The idea is to optimize the weaker GLB in (\ref{simplified_GLB}). By relating the quadratic minimization solution of $\mathcal{F}_{K,M,N,m}(\mathbf{w})$ to the Chebyshev polynomials of the second kind, one can obtain the weight vector\footnote{Although it looks similar to that in [\ref{LiuParaGuanBozas14}, Lemma 2], such a weight vector is more generic as it applies to QCSS with different $M\geq 2$.} below.

Let $K\leq MN^2$ and $\cos \varphi=1-\frac{K}{MN^2}$. Also, let $m$ be an even positive integer with $m \varphi<\pi+\varphi$.
For $\varphi_0=(\pi-m\varphi+\varphi)/2$, define the following weight vector
\begin{equation}\label{LevWtVec_equ}
{w}_i=\left \{
\begin{array}{ll}
\frac{\sin \frac{\varphi}{2}}{\sin \frac{m\varphi}{2}} \sin (\varphi_0+i\varphi), & i\in \{0,1,\cdots,m-1\};\\
0,   & i\in \{m,m+1,\cdots,2N-2\}.
\end{array}
\right .
\end{equation}

Setting $m=\left \lfloor \frac{\pi}{\varphi}\right \rfloor+1$, one can minimize $\mathcal{F}_{K,M,N,m}(\mathbf{w})$ in (\ref{mini_F}) over different $m$ and get a generalized version of the Levenshtein bound in [\ref{Levenshtein99}, \textit{Corollary 4}] as follows.

\vspace{0.1in}

\begin{corollary}\label{coro_bd1}
\begin{equation}\label{lev_bd_cor4}
\delta^2_{\max} \geq M\left (N-\left \lceil \frac{\pi N}{\sqrt{8K/M}} \right \rceil \right ),~~\text{for}~K\leq MN^2.
\end{equation}
\end{corollary}

\vspace{0.1in}


As $N\rightarrow\infty$, the lower bound in (\ref{lev_bd_cor4}) is tighter than the Welch bound in (\ref{Welch_bound_for_cc}) if and only if
\begin{equation}\label{equ_coro2_tighter}
\frac{1}{2}+\frac{M}{2K}> \lim_{N\rightarrow\infty} \frac{1}{N} \left \lceil \frac{\pi N}{\sqrt{8K/M}} \right \rceil  = \frac{\pi}{\sqrt{8K/M}},
\end{equation}
or equivalently,
\begin{equation}\label{equ_coro2_tighter2}
\frac{K}{M}>\frac{\pi^2}{4}-1+
{\sqrt{\left (\frac{\pi^2}{8}-1 \right )\frac{\pi^2}{2}}} \approx 2.541303,
\end{equation}
where the right-hand side of (\ref{equ_coro2_tighter}) is obtained from (\ref{lev_bd_cor4}).

Recall that as $N\rightarrow\infty$, a necessary condition (cf. \textit{Remark \ref{rmk_nece_cond}}) for the GLB to be tighter than the corresponding Welch bound is
\begin{equation}\label{nece_cond_QCSSBd2_2}
\frac{K}{M} \geq
\frac{\lim\limits_{N\rightarrow\infty}\overline{K}+1}{M}
=\frac{\left \lfloor\frac{\pi^2M}{4}\right \rfloor+1}{M}.
\end{equation}
Clearly,
\begin{equation}\label{overlineK}
\frac{\lim\limits_{N\rightarrow\infty}\overline{K}}{M}
=\frac{\left \lfloor\frac{\pi^2M}{4}\right \rfloor}{M} \leq \frac{\pi^2}{4} \approx 2.467401,
\end{equation}
which is smaller than the right-hand side of (\ref{equ_coro2_tighter2}).

It can be asserted that the resultant GLB obtained from the weight vector in (\ref{LevWtVec_equ}) with $m=\left \lfloor {\pi}/{\varphi}\right \rfloor+1$ is tighter if and only if the value of $M$ satisfies the condition
\begin{equation}\label{ch5_dM_equ2}
d_3(M)  \triangleq \frac{\left \lfloor \frac{\pi^2M}{4}\right \rfloor+1}{M}-\left [ \frac{\pi^2}{4}-1+
{\sqrt{\left (\frac{\pi^2}{8}-1 \right )\frac{\pi^2}{2}}}\right ]  > 0.
\end{equation}
This is because when condition (\ref{ch5_dM_equ2}) is satisfied,
$K\geq \lim\limits_{N\rightarrow\infty} \overline{K}+1$ is not only a necessary condition [cf. (\ref{nece_cond_QCSSBd2_2})] but also a sufficient condition [cf. (\ref{equ_coro2_tighter2})] for the GLB to be asymptotically tighter than the Welch bound.


In Fig. \ref{fig_dM}, $d_3(M)$ versus $M$ is plotted. By identifying $M$ satisfying $d_3(M)>0$ [shown in (\ref{ch5_dM_equ2})],  we have the following theorem.


\vspace{0.1in}
\begin{theorem}\label{tighter_M_rmk2}
The GLB in (\ref{lev_bd_cor4}) which arises from the weight vector in (\ref{LevWtVec_equ}) is asymptotically tighter than the Welch bound for \textit{all} $K\geq \overline{K}+1$ if and only if
\begin{equation}\label{distri_optimal_M2}
M \in  \Bigl \{3,5,7,9 ,11\Bigl \}.
\end{equation}
\end{theorem}

\vspace{0.1in}

\subsection{Discussions}

Denote by $\mathrm{B}_1,\mathrm{B}_2,\mathrm{B}_3$ the optimized asymptotic lower bounds in (\ref{aym_lwrbd_wec3_equ}), (\ref{aym_lwrbd_wec2_equ}), (\ref{lev_bd_cor4}),
respectively. We remark that (1), Both $\mathrm{B}_1$ and $\mathrm{B}_2$ are greater than $\mathrm{B}_3$ for any $M\geq2$; (2), $\mathrm{B}_1>\mathrm{B}_2$ except for $M\in \{3,5,7,9\}$. The proof is omitted as it can be easily obtained from the tightness analysis in Section III-B and Section IV.

To further visualize their relative strengths of these three lower bounds, we calculate in Table I the ratio values of $\frac{\mathrm{B}_1}{\mathrm{B}_{\text{W}}},\frac{\mathrm{B}_2}{\mathrm{B}_{\text{W}}},\frac{\mathrm{B}_3}{\mathrm{B}_{\text{W}}}$ with $M\in \{2,3,\cdots,25\}$, where $N=2048,K=\overline{K}+1$ and $\mathrm{B}_{\text{W}}$ denotes the corresponding Welch bound. A ratio value which is larger than 1 corresponds to a tighter GLB (over the Welch bound). With Table I, one may verify the three sets of $M$ for tighter GLB in \textit{Theorems 1-3} as well as the above-mentioned remark in this subsection. In particular, we can see that $\frac{\mathrm{B}_1}{\mathrm{B}_{\text{W}}}>1$ for all $M\geq2$, showing that weight vector 1 is superior than the other two as it is capable of tightening the GLB for all possible $M$, asymptotically.

\begin{table*}[!ht]
\setlength{\tabcolsep}{4pt}
\centering \label{table_of_small_value}
\caption{Comparison of $\frac{\mathrm{B}_1}{\mathrm{B}_{\text{W}}},\frac{\mathrm{B}_2}{\mathrm{B}_{\text{W}}},\frac{\mathrm{B}_3}{\mathrm{B}_{\text{W}}}$ with different $M$, where $N=2048,K=\overline{K}+1$. }
\begin{tabular}{|c||c|c|c|c|c|c|c|c|c|c|c|c|c|}
\hline
$M$ & 2 & 3 & 4 & 5 & 6 & 7 & 8 & 9 & 10  & 11 & 12 & 13\\
 \hline \hline
$\frac{\mathrm{B}_1}{\mathrm{B}_{\text{W}}}$ & 1.0043  &  1.0241  &  1.0043  &  1.0166  &  1.0042 &   1.0133  &  1.0042 &   1.0113  &  1.0042  & 1.0101 & 1.0042 & 1.0092
\\ \hline
$\frac{\mathrm{B}_2}{\mathrm{B}_{\text{W}}}$ & 1.0026  &  1.0293  &  1.0025  &  1.0187  &  1.0025 &   1.0141  &  1.0025 &   1.0116  &  1.0025  &  1.0099 & 1.0025  &  1.0088
\\ \hline
$\frac{\mathrm{B}_3}{\mathrm{B}_{\text{W}}}$ & \textit{0.9909}  &  1.0232  &  \textit{0.9909}  &  1.0106  &  \textit{0.9910} &   1.0049  &  \textit{0.9910} &   1.0025  &  0.9910  &  1.0002 & \textit{0.9910}  &  \textit{0.9988}
 \\ \hline \hline
$M$ & 14 & 15 & 16 & 17 & 18 & 19 & 20 & 21 & 22  & 23 & 24 & 25\\
 \hline \hline
$\frac{\mathrm{B}_1}{\mathrm{B}_{\text{W}}}$ & 1.0042 &   1.0086 &   1.0042 &  1.0003 &  1.0042  &  1.0007 &  1.0042 &  1.0010  &  1.0042 &  1.0013  &  1.0042 &   1.0016
\\ \hline
$\frac{\mathrm{B}_2}{\mathrm{B}_{\text{W}}}$ & 1.0025 &  1.0079  &  1.0025  & \textit{0.9977}  &  1.0025  &  \textit{0.9983} &  1.0025 &  \textit{0.9987}  &  1.0025 &  \textit{0.9990}  &  1.0025 &   \textit{0.9993}
\\ \hline
$\frac{\mathrm{B}_3}{\mathrm{B}_{\text{W}}}$ & \textit{0.9910} &  \textit{0.9969}  &  \textit{0.9910}  &  \textit{0.9841} &  \textit{0.9910}  & \textit{0.9849}  &  \textit{0.9910} &  \textit{0.9859}  &  \textit{0.9910} &  \textit{0.9870}  &  \textit{0.9910} &  \textit{0.9865}
 \\ \hline
\end{tabular}
\end{table*}

\section{Conclusions}

The generalized Levenshtein bound (GLB) in [\ref{LiuGuanMow14}, \textit{Theorem 1}] is an aperiodic correlation lower bound for quasi-complementary sequence sets (QCSSs) with \textit{number of channels} not less than 2 (i.e., $M\geq2$). Although GLB was shown to be tighter than the corresponding Welch bound [i.e., (\ref{Welch_bound_for_cc})] for certain cases, there exists an ambiguous zone [shown in (\ref{GLBgap_}) and (\ref{GLBgap})] in which the tightness of GLB over Welch bound is unknown. Motivated by this, we aim at finding a properly selected weight vector in the bounding equation for a tighter GLB for \textit{all} (other than \textit{some}) $K\geq \overline{K}+1$, where $K$ denotes the set size, and $\overline{K}$ is a value depending on $M$ and $N$ (the sequence length). As the GLB is in general a non-convex fractional quadratic function of the weight vector, the derivation of an analytical solution for a tighter GLB for \textit{all} possible cases is a challenging task.

The most significant finding of this paper is weight vector 1 in (\ref{wgtvec3_equ}) which is obtained from a frequency-domain optimization approach. We have shown that its resultant GLB in (\ref{coro_GLB_from_wetvec3-equ}) is tighter than Welch bound for \textit{all} $K\geq \overline{K}+1$ and for \textit{all} $M\geq2$, asymptotically. This finding is interesting as it explicitly shows that the GLB tighter condition given in [\ref{LiuGuanMow14}, \textit{Theorem 2}] is not only necessary but also sufficient, asymptotically, as shown in \textit{Theorem \ref{th4OptWtVec}}. Interestingly, we have proved in Section III-C that weight vector 1 in (\ref{wgtvec3_equ}) is local minimizer of the GLB under certain asymptotic conditions.

We have shown that both weight vectors 2 and 3 [given in (\ref{sine_shape_weight_vector}) and (\ref{LevWtVec_equ}), respectively] lead to tighter GLBs for \textit{all} $K\geq \overline{K}+1$ but only for certain small values of $M$ not less than 2. Note that although they were proposed in \cite{LiuParaGuanBozas14}, the focus of \cite{LiuParaGuanBozas14} was on the tightening of Levenshtein bound for traditional single-channel (i.e., $M=1$) sequence sets, whereas in this paper we have extended their tightening capability to GLB for multi-channel (i.e., $M\geq2$) QCSS. Furthermore, we have shown in \textit{Theorem \ref{tighter_M_rmk}} and \textit{Theorem \ref{tighter_M_rmk2}} that weight vector 2 is superior as its admissible set of $M$ [see (\ref{distri_optimal_M})] is larger and subsumes that of weight vector 3. 

\end{document}